# Simulation of the Efficiency of CdS/CIGS Tandem Multi-Junction Solar Cells Using AMPS-1D

Ashrafalsadat S. Mirkamali[1,2], Khikmat Kh. Muminov[1]

[1]*S.U.Umarov Physical-Technical Institute, Academy of Sciences of the Republic of Tajikistan, 299/1 Aini Ave, Dushanbe 734062, Tajikistan*
*e-mail: khikmat@inbox.ru*
[2]*Permanent address: Department of Science and Engineering, Behshahr Branch, Islamic Azad University, Behshahr, Iran*
*e-mail: ash.mirkamali@gmail.com*

## Abstarct

In this paper we conduct numerical simulation of CdS/CIGS solar cells by use of the AMPS-1D software aiming to formulate the optimal design of the new multi-junction tandem solar cell providing its most efficient operation. We start with the numerical simulation of single-junction CdS/CIGS solar cells, which shows that its highest efficiency of 17.3% could be achieved by the thickness of CIGS p-layer of 200 nm. This result is in a good agreement with experimental data where the highest efficiency was 17.1% with the solar cell thickness of 1 micron. By use of the results of the numerical simulation of the single-junction solar cells we developed the design and conducted optimization of the new multi-junction tandem CdS/CIGS solar cell structure. Numerical simulation shows that the maximum efficiency of this solar cell is equal to 48.3%, which could be obtained with the thickness of the CIGS p-layer of 600 nm at a standard illumination of AM 1.5.

# Introduction

In this paper, as an object of study we consider solar cells of the second generation copper gallium indium diselenide CIGS, and investigate the influence of the thickness of the semiconductor layers on the output parameters of the solar cell, such as a short circuit current density $J_{sc}$, the open circuit voltage $V_{oc}$, the fill factor FF and the efficiency of conversion EFF, by means of numerical simulation using the AMPS-1D program. Then, in order to obtain the maximum efficiency, taking into account the results obtained for single-junction solar cells new tandem multilayer structure consisting of layers of two solar cells connected with each other back to back are designed and studied.

It should be noted that for the solar cells made on the basis of such perspective material as copper indium gallium diselenide (CIGS), experimentally the conversion efficiency of 9% [1] and 12.8% [2], are expected for ultrathin solar cells grown by close space vapor transport (CSVT) and coevaporation methods, respectively. In order to optimize the structure of the thin-film CIGS solar cells in [3-4] numerical studies were conducted. In particular, the type of substrate [3-4] has been investigated in detail; however, it is worth to note that while there are few studies of ultrathin CIGS solar cells. A comprehensive analysis carried out in [5] has shown that the best performance is achieved when the thickness of the CIGS absorber p-layer varies within the 0.2 - 0.3 micron, when the carrier density in the absorbing layer ranges from $10^{12}$ to $10^{16}$cm$^{-3}$. Formation of the absorbing layer of n-type plays an important role in optimizing the CIGS ultrathin solar cells, although improving of the conversion efficiency can be expected by increasing the number of charge carriers over $10^{18}$ cm$^{-3}$ in the buffer layer $In_2Se_3$. The conversion efficiency of over 15% can be obtained in the maximum quantum efficiency of 80% [6].

In this paper we study single-junction solar cells made on the basis of $Cu(In_xGa_{1-x})(Se_xS_{1-x})$ or CIGS, which is a four-component alloy of elements of I-III-IV groups. By use of numerical simulation we show that the optimum thickness of the absorbing CIGS p-layer is 200 nm at the conversion efficiency equal to 17.3%. It was found that for this solar cell among all the major parameters only the density of the short-circuit current has a strong dependence on the thickness of the CIGS p-layer.

Further we develop a new design of optimal multi-junction CdS/CIGS tandem solar cell and by numerical simulation show that the highest efficiency of this solar cell in 48.3% could be obtained with the thickness of CIGS p-layer of 600nm.

## Simulation of CdS/CIGS thin film solar cells

$Cu(In_xGa_{1-x})(Se_xS_{1-x})$ or CIGS is a four-component alloy of I-III-IV groups, which is formed by replacing the atoms of indium by gallium atoms in the sublattice of $CuInSe_2$ (CIS), which crystallizes into the stable structure of chalcopyrite [7]. The purpose of replacing the anion or cation is to change the width of CIS bandgap (1.02 eV), so as to keep it in the optimal range for the photoelectric conversion. The electronic conductivity of these materials could be explained in terms of chemistry of their internal defects.

Vacancies of copper and indium (i.e. excess of selenium) lead to semiconductor material of p-type having a carrier density ranging from 0.15 to $2 \times 10^{17}$ $cm^{-3}$, while selenium vacancies lead to n-type conductivity. The most important defect for carrier recombination is InCu (In in opposite site to Cu), both as a result of its low energy of formation, and because of its predicted level of 0.34 eV, which is below the conduction band maximum (CBM), lead to substantial compensation of p-type material by reducing its activity through the formation of [2VCu + InCu] defect complexes [8].

CIGS photovoltaic device could be obtained by forming of p-n-heterojunctions on the CdS thin film, as shown in Fig. 1. Like the CdTe devices the role of CdS semiconductor of n-type, whose bandgap of 2.4 eV, is not only to form a p-n-junction with the absorber, but also to serve as a window layer that transmits incident light with relatively low losses on the absorption and reflection. Manufacturing usually begins with the deposition of the back contact, which is made of Mo, followed by a p-type absorber - a thin window layer of CdS (50-100 nm), and ZnO doped with Al is introduced as a transparent front contact. The conductive oxide ITO layer may also be added to this contact in order to maximize the absorption and, in turn, the current density obtained from these solar cells. The advantage of CIGS solar cell lies in the flexible substrates (such as soda-lime glass, aluminum foil, or high temperature polyamide) to which these layers may be deposited [9].

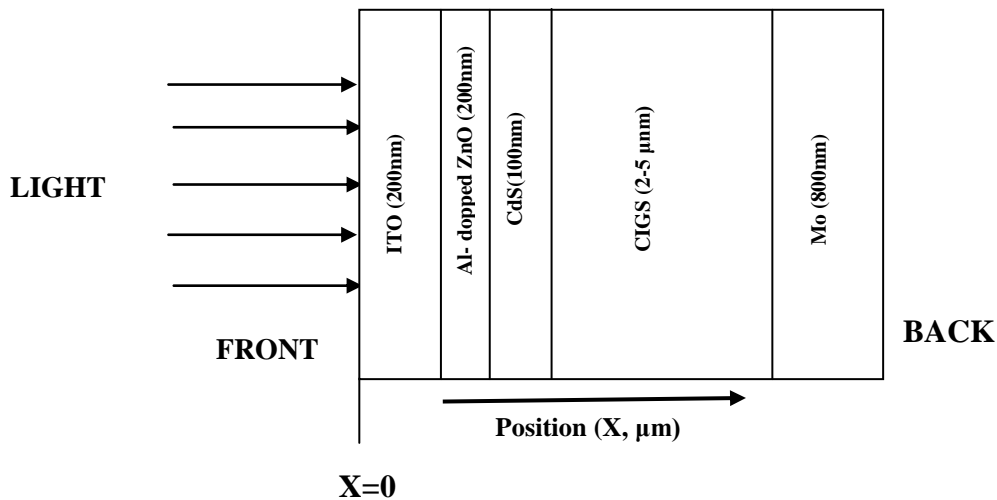

**Figure 1.** A typical structure of the Cu-In-Ga-S, Se (CIGS) solar cell [7].

CIGS possesses a direct optical band, and it is usually grown on a substrate of soda lime glass or flexible substrates. The advantage of using a CIGS as a solar cell material is the ability to create devices with thin absorbent layers on various substrates, which leads to a significant reduction of production costs and reduce production time. In a standard device an absorber thickness usually is 2 microns. Decreasing the thickness on each 0.5 microns can save up to 75% consumption of semiconductor materials and the corresponding deposition time is also reduced four times. For example, reduce the thickness of the deposited layer from 1.8 microns to 0.15 microns [10, 11] results in a reduction of deposition time from one hour to 8 minutes. In in the paper [12] it has been experimentally shown that the efficiency of 13.1% could be achieved when the thickness of the CIGS solar cell is 0.86 microns. At the same time, in the paper [13] it was shown that when the device thickness of 1 micron the efficiency is 9.9%. It means that with decreasing of the solar cell thickness the efficiency increase is observed. Thinnest element with acceptable performance [14] had a thickness of 150 nm and 5% efficiency. For a CIGS solar cell with 1 micron thickness the highest registered efficiency of conversion was 17.1% [15].

The concept of this solar cell is the use of three layers: a ZnO n-layer (with a band gap of 3.30 eV), which is used as a transparent contact layer, a CdS n-layer (2.40 eV), which is a frontal absorbing layer, and a CIGS p-layer (1.15 eV) - absorbing layer of p-type. This configuration is very popular for CIGS based devices. The parameters used in the numerical simulation of the device are shown in the Table 1.

The purpose of the numerical simulation of CIGS solar cells is to optimize the design of solar cell with reduced thickness, which was varied from 30 nm to 1000 nm, as shown in Fig. 2. This graph shows that the efficiency of the solar cell increases by increasing the thickness of the CIGS absorbing p-layer, and then decreases gradually. The optimum thickness of the absorbing CIGS p-layer was 200 nm, at the efficacy of 17.3%. As one can see from the output data of the simulation among all the solar cell parameters only the density of the short-circuit current has a strong dependence on the thickness of the CIGS p-layer. With increasing thickness of the absorber, it is reduced. Comparison of the simulation results and experimental data shows that the current decline of the experimental data is much steeper than predicted by the simulation. This difference is due to the processing of grain boundaries of the layers [16, 17].

Thus, the numerical simulation shows that for CdS/CIGS solar cell the maximal efficiency is 17.3% at the optimum thickness of the absorbing CIGS p-layer equal to 200 nm.

For the similar solar cell the highest registered efficiency was 17.1% at the solar cell thickness of 1 micron (A.O.Pudov et al. [15]), which is in good agreement with our results.

**Table 1.** Baseline CIGS parameters used in the numerical simulation

| Material | Band gap (eV) | Conductivity type | Conduction Band | Valence Band | Electron Affinity (eV) | Electron Mobility (cm2 /v/s) | Hole Mobility (cm2 /v/s) | Free Carrier Concentration (cm$^{-3}$) | Relative Permittivity |
|---|---|---|---|---|---|---|---|---|---|
| **ZnO** | 3.30 | N | $2.2*10^{18}$ | $1.8*10^{19}$ | 4.10 | 100.0 | 25.0 | $1.0*10^{18}$ | 9.0 |
| **CdS** | 2.40 | N | $2.2*10^{18}$ | $1.8*10^{19}$ | 4.0 | 100.0 | 25.0 | $1.1*10^{18}$ | 10.0 |
| **CIGS** | 1.15 | P | $2.2*10^{18}$ | $1.8*10^{19}$ | 4.5 | 100.0 | 25.0 | $2.0*10^{16}$ | 13.4 |

**Figure 2.** The output data of the solar cell (short-circuit current density $J_{sc}$, open-circuit voltage $V_{oc}$, fill factor FF and efficiency EFF) depending on the thickness of the CIGS p-layer

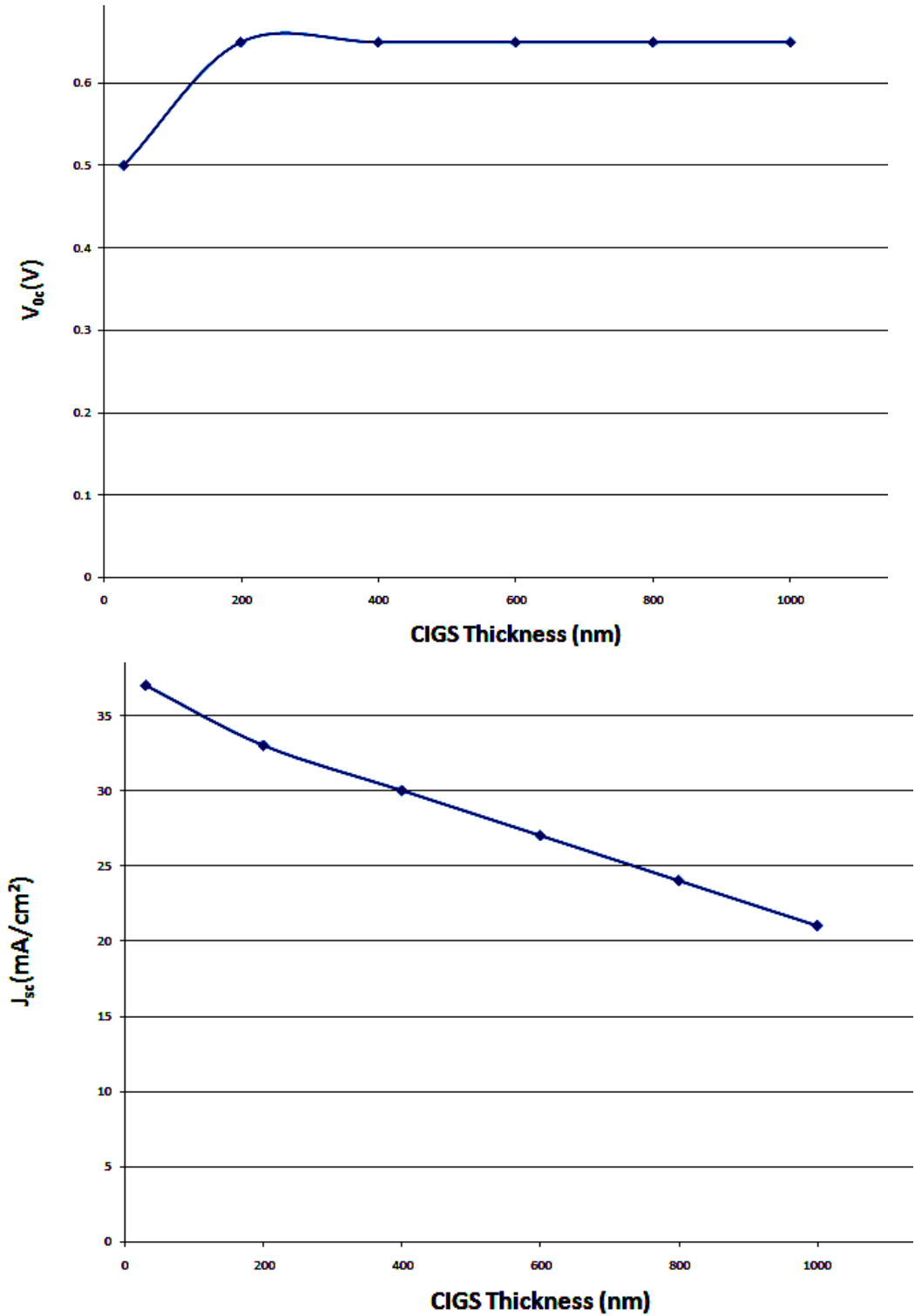

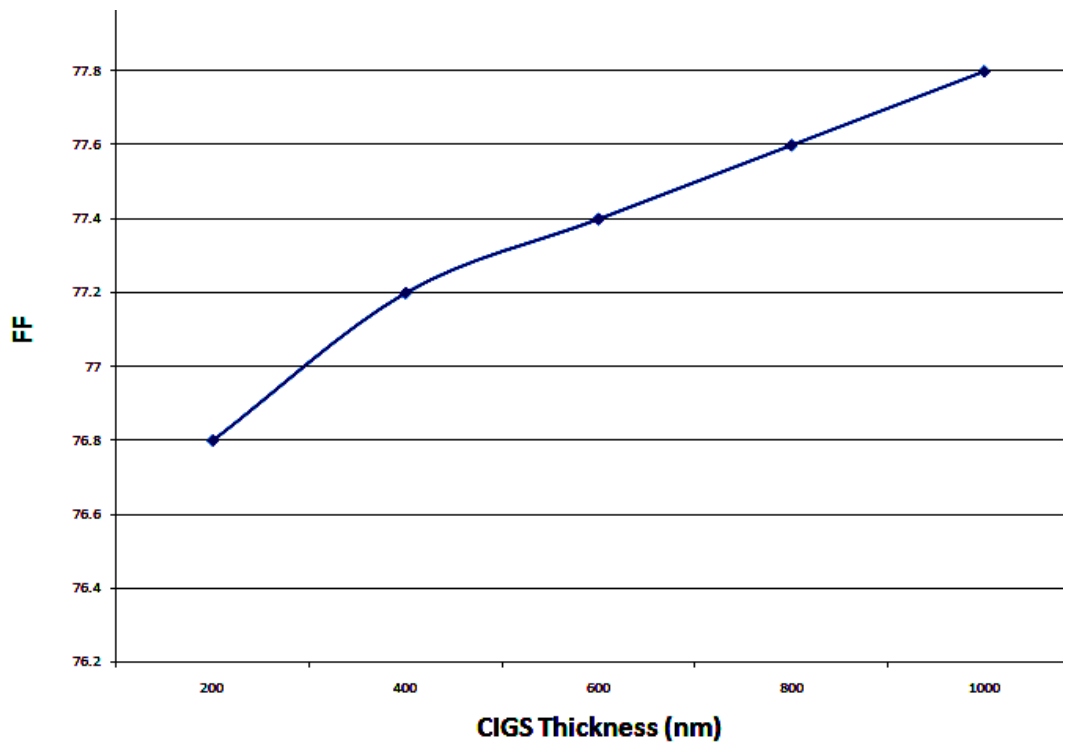

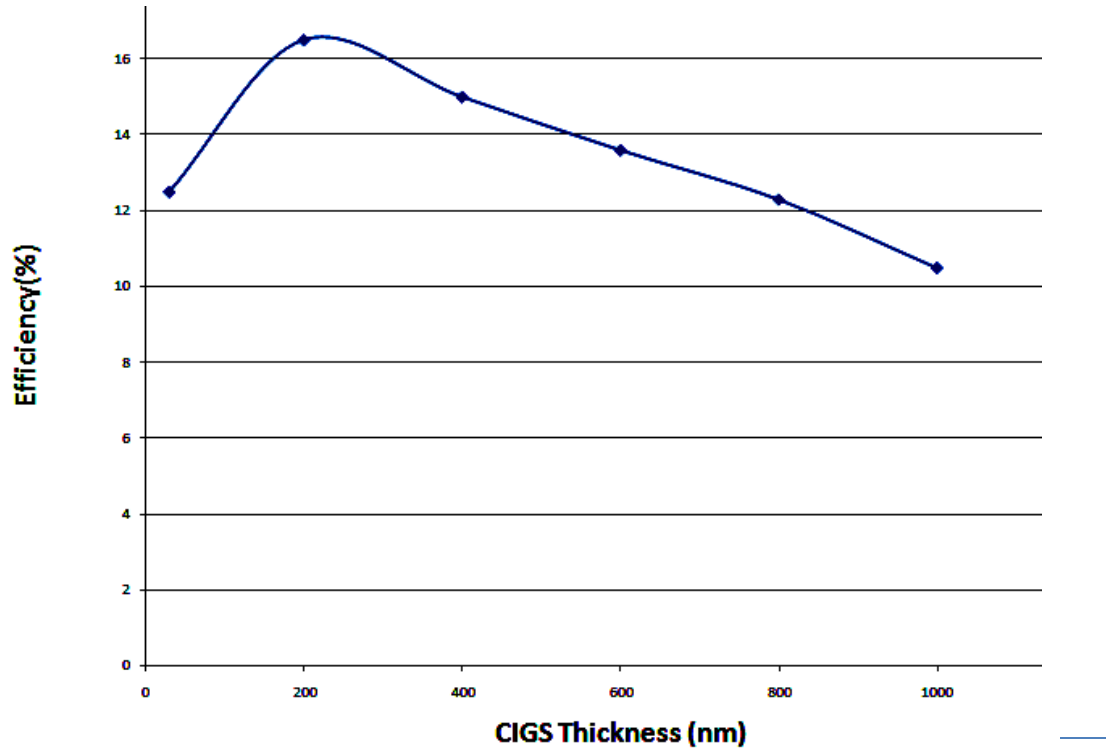

# Simulation of tandem multi-junction CdS/CIGS solar cells

Multijunction solar cells consist of several p-n-junctions of different semiconductor materials with different band gaps, which have the ability to absorb a large part of the solar energy spectrum. Multijunction solar cells on the basis of elements of III-V groups represent the new technology as compared with single-junction solar cells, providing in the same conditions, nearly twice the efficiency [18], but the production costs of these elements are very high, and they used only in special cases, for example in space technologies.

Another way to increase the efficiency of solar cells is the use of such cells, each of which uses a certain part of the spectrum of solar radiation for the production of electric current. Tandem solar cells can be used as single or serial connection, where the current in both cases is similar. The structure of the solar cell in the series connection is very simple, but because of limitations caused by wide bandgap, solar cells of a single compound, from the viewpoint of efficiency, are more optimal.

The most common method of creating tandem solar cells is their growing even when the layers are built up sequentially on the substrate and provide a tunneling contact of layers in individual cells. Due to increasing of number of band gaps, the efficiency of the cell also icreases. The upper part of cell has the largest width of band gap; it absorbs photons which have higher energy of the spectrum of incident light, while the lower part of the cell has a small band gap width, and hence, provides the absorption of the low energy photons [19].

This section presents the research, aimed at increasing the efficiency of solar cells and the design of the solar cell structure having the most optimal efficiency, using the results of investigations of single-junction solar cells whose efficiency is studied in the previous section. By use of AMPS-1D software we conduct a study of new multijunction tandem CdS/CIGS solar cell structures in order to determine the most optimal one with the highest efficiency.

In order to enhance the effeticiency and to determine the most optimal solar cell we use two single-contact solar cells connected back to back. In Fig. 3 a proposed solar cell layers and the order of their arrangement is depicted. This multilayer solar cell is made of two CdS layers of p-type and n-type and two layers of CIGS of p-type and n-type. On the top of the solar cell the layer of indium tin oxide (ITO) of 200 nm thick is located to provide greater absorption of light flux by the solar cell and the lower layer is a molybdenum one 500 nm thick for reflecting the light flux.

| ITO (200нm) |
|:---:|
| p- C IGS (60нm) |
| n- CIGS (100нm) |
| p- CdS (400нm) |
| n- CdS (1000нm) |
| Mo (500нm) |

**Figure 3.** Schematic representation of the CdS/CIGS multijunction tandem solar cell

The parameters of each layer of the solar cell used as input data in the numerical simulation by AMPS-1D are shown in the Table 2. In this simulation the thickness of the CdS p-layer is chosen equal to 50 nm and the CdS n-layer is 200 nm.

As a result of our previous numerical simulation we know that the most optimal solar cell should have a thickness of CIGS n-layer equal to 3000 nm. In our simulation we kept it constant and varied the thickness of CIGS p-layer in the range from 400 nm to 2000 nm. Upper buffer layer serves to provide greater absorption of the blue part of the solar spectrum. The photons, which are not absorbed in the upper layers, would be absorbed in the lower absorbent layer and produce electron-hole pairs. As a result, the overall efficiency of the solar cell, which is the total efficiency of the upper and lower elements, is increased.

The graphs (Fig. 4) show the dependence of the open circuit voltage $V_{oc}$, short-circuit current $J_{sc}$, fill factor FF and efficiency ŋ on the thickness of CIGS p-layer, which we have obtained in numerical experiments. The results show that the most optimal solar cell with the thickness of CIGS p-layer of 600 nm at a standard sunlight of AM 1.5 is the element with the highest efficiency equal to 48.3%.

This efficiency is more than twice of the efficiency of CdS/CIGS single-junction solar cell under similar conditions.

**Table 2.** The parameters of CdS/CIGS multijunction solar cell

| Material | Band gap (eV) | Conductivity type | Conduction Band | Valence Band | Electron Affinity (eV) | Electron Mobility (cm2 /v/s) | Hole Mobility (cm2 /v/s) | Free Carrier Concentration (cm$^{-3}$) | Relative Permittivity |
|---|---|---|---|---|---|---|---|---|---|
| ZnO | 3.30 | N | 2.2*1018 | 1.8*1019 | 4.10 | 100.0 | 25.0 | 1.0*1018 | 9.0 |
| CdS | 2.40 | N | 2.2*1018 | 1.8*1019 | 4.0 | 100.0 | 25.0 | 1.1*1018 | 10.0 |
| CdS | 2.40 | p | 2.2*1018 | 1.8*1019 | 4.0 | 100.0 | 25.0 | 1.1*1016 | 10.0 |
| CIGS | 1.15 | P | 2.2*1018 | 1.8*1019 | 4.5 | 100.0 | 25.0 | 2.0*1014 | 13.4 |
| CIGS | 1.15 | N | 2.2*1018 | 1.8*1019 | 4.5 | 100.0 | 25.0 | 2.0*1016 | 13.4 |

**Figure 4.** The dependence of the output parameters (short circuit current density $J_{sc}$, open-circuit voltage $V_{oc}$, the fill factor FF and the efficiency EFF) of CdS/CIGS multijunction tandem solar cell on the thickness of the CIGS p-layer

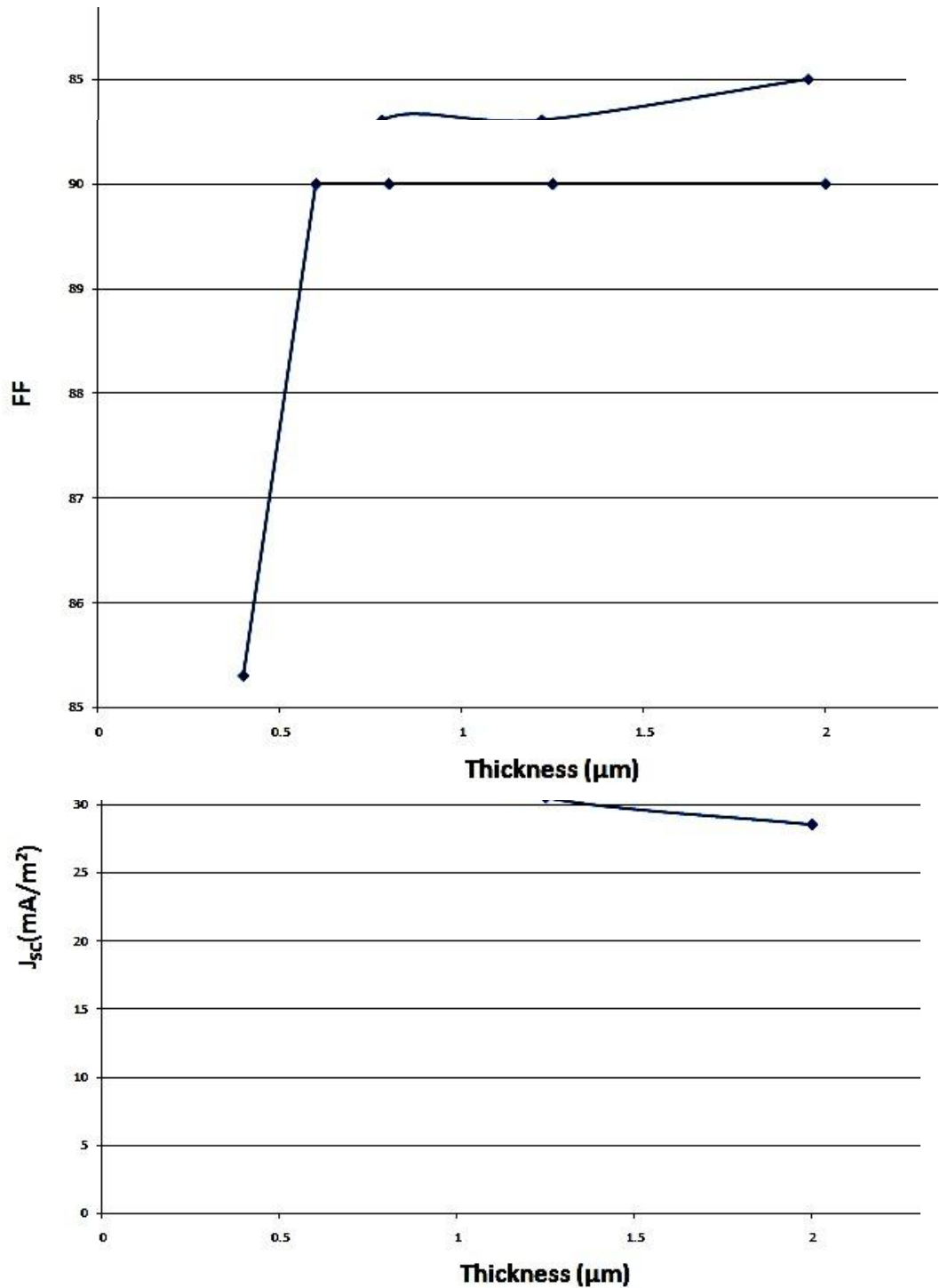

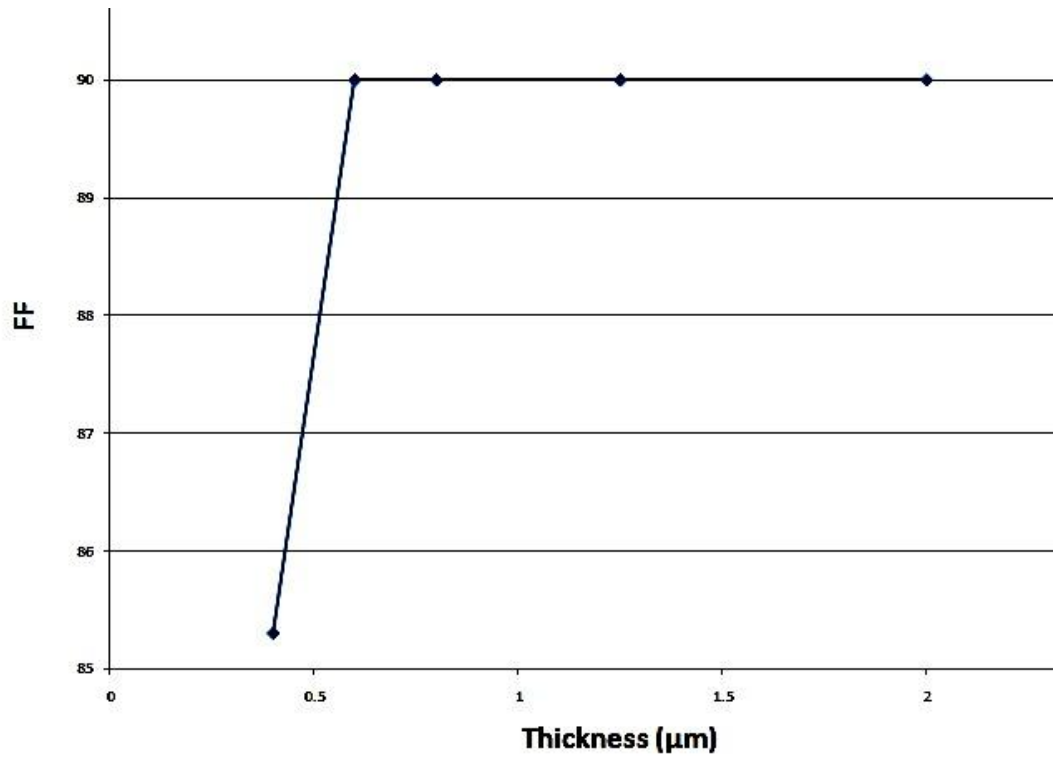
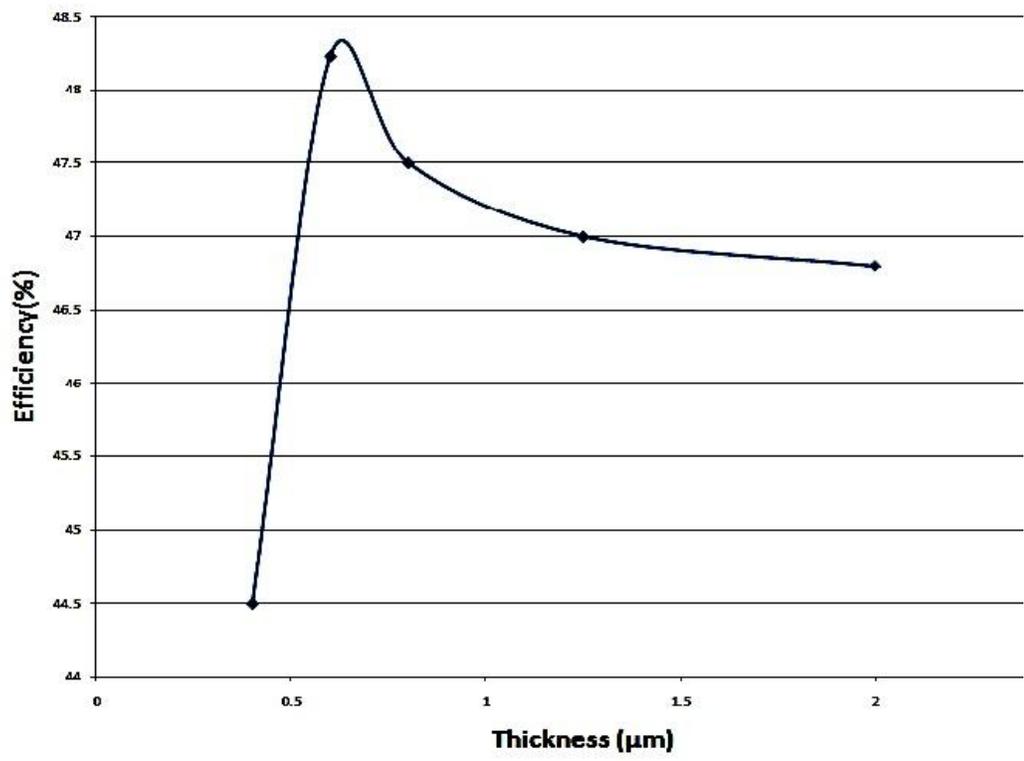

## Conclusion

Our numerical study of single-junction CdS/CIGS solar cells shows that its highest efficiency of 17.3% could be achieved by the thickness of CIGS p-layer of 200 nm. This result is in a good agreement with experimental data obtained for the same solar cell by A.O.Pudov et al. [15], wherein the highest efficiency was 17.1% with the solar cell thickness of 1 micron.

By use of the results of the numerical simulation of the single-junction solar cells we developed the design and conducted optimization of the new multi-junction tandem CdS/CIGS solar cell structure. Numerical simulation shows that the maximum efficiency of this solar cell is equal to 48.3% with the thickness of the CIGS p-layer of 600 nm at a standard illumination of AM 1.5.

Thus, numerical simulation based on the use of one-dimensional analysis of microelectronic and photonic structures (AMPS-1D) software for the analysis of copper indium gallium diselenide (CIGS) solar cells allowed us to formulate the optimal design of the new multi-junction tandem solar cell providing its most efficient operation.

# References

[1] Djessas K., Yapi S., Masse G., Ibannain M., Gauffier J.L. Diffusion of Cu, In and Ga in $In_2Se_3/CuGaSe_2/SnO_2$ thin film photovoltaic structure//Journal of Applied Physics, 2004.- V.95.- N.8.- P.4111-4116.

[2] Nakada T., Hirabayashi Y., Tokado T., Ohmori D., Mise T. Novel device structure for Cu (In,Ga)$Se_2$ thin film solar cells using transparent conducting oxide back and front contacts // Solar Energy,2004.- V. 77.- N.6.-P. 739 -747.

[3] Bouloufa A., Djessas K., Zegadi A. Numerical simulation of $CuInxGa1_{-x}Se_2$ solar cells by AMPS-1D// Thin Solid Films, 2007.-V.515.-N.15.- P. 6285-6287.

[4] Alagappan S.A., Mitra S. Optimizing the design of CIGS-based solar cells: a computational approach // Materials Science and Engineering:B,2005.- V. 116 .-N.3.- P. 293-296.

[5] Bouchama I., Djessas K., Djahli F., Bouloufa A.Simulation approach for studying the performances of original superstrate CIGS thin films solar cells// Thin Solid Films, 2011.-V. 519 .- P. 7280–7283.

[6] S.M.Sze.photonic Devices'in semiconductor Devices-Physics and technology/ S.M.Sze.-2$^{nd}$ ed.-U.S.A.: John Wiley and sons , 2002.- P.320-330.

[7] Birkmire R.W., Eser E. Polycrystalline Thin Film Solar Cells: Present Status and Future Potential// Annu. Rev. Mater. Sci, 1997.-V. 27.- P.625-653.

[8] Choi S.G., Zhao H. Y., Persson C., Perkins C. L., Donohue A. L., To B., Norman A. G, Li J., Repins I. L. Dielectric function spectra and critical-point energies of $Cu_2ZnSnSe_4$ from 0.5 to 9.0 eV// Journal Of Applied Physics, 2012.-V. 111.-P. 033506.1- 033506-6.

[9] Wolden C.A., Kurtin J., Baxter J.B., Repins I., Shaheen S.E., Torvik J.T., Rockett A.A., Fthenakis V.M., Aydil E.S. Photovoltaic Manufacturing: Present status, future prospects, and research needs/ J. Vac. Sci. Technol., 2011.-V. 29 .-N.3.- P.1-16.

[10] Contreras M.A., Egaas B., Ramanathan K., Hiltner J., Swartzlander A., Hasoon F., Noufi R. Progress Toward 20% Efficiency in Cu(In,Ga) Se2 Polycrystalline Thin-Film Solar Cells // Prog.in PV,199.-V.7.-p- 311-316.

[11] Lundberg O., Edoff M., Stolt L. The effect of Ga-grading in CIGS thin film solar cells // Thin Solid Films,2004.- V.480.- P.520-525.

[12] Sites J.R., Characterization and Analysis of CIGS and CdTe Solar Cells// NREL Technical Monitor: Bolko von Roedern,2009.- 55 pages.

[13] Yamaguchi M., Toyota T.I, Nagoya J., Luque A. High efficiency and high concentration in photovoltaics // Electron Devices, IEEE Transactions,1999.-V. 46.-N.10.- P. 2139 – 2144.

[14] Nanu M., Schoonman J., Goossens A. Nanocomposite Three-Dimensional Solar Cells Obtained by Chemical Spray Deposition // Nano Lett., 2005.-V. **5** .-N.9.-P. 1716–1719.

[15] Pudov A.O., Contreras M.A., Nakada T., Schock H.W. CIGS J-V Distortions in the Absence of Blue Photons //Thin Solid Films, 2005.- V. 480-481.-P. 273-278.

[16] Hergutha A., Horbelta R., Wilkinga S., Jobb R., Hahna G. Comparison of BO Regeneration dynamics in PERC and Al-BSF solar cells// Energy Procedia, 2015.-V. 77 .-P. 75 – 82.

[17] Morozov O.I. Structure of Symmetry Groups via Cartan's Method: Survey of Four Approaches// Symmetry, Integrability and Geometry: Methods and Applications, 2005.- V.1.- P.1-14.

[18] Luque A., Martı́ A., Stanley C., Lo´pez N., Cuadra L., Zhou D., Mc Kee A. General equivalent circuit for intermediate band devices: Potentials, currents and electroluminescence// J. Appl. Phys..-V. 96.-N.03.-p. 903-909.

[19] http://www.pveducation.org/pvcdrom/solar-cell-operation/tandem-cells